\documentclass[12pt,a4paper,fleqn,twoside]{article}

\usepackage{amsmath}
\usepackage{amssymb}

\newtheorem{theorem}{Theorem}[section]
\newtheorem{lemma}[theorem]{Lemma}

\newtheorem{example}[theorem]{Example}

\newenvironment{proof*}{\paragraph{Proof.}}{}

\newcommand{\arrow}{\rightarrow}
\newcommand{\map}{\mapsto}

\newcommand{\bb}[1]{\mathbb{#1}}

\newcommand{\alg}{\mathfrak{g}}

\newcommand{\pr}{\partial}

\newcommand{\me}{\geqslant}
\newcommand{\les}{\leqslant}

\newcommand{\deriv}[2]{\frac{\partial #1}{\partial #2}}
\newcommand{\bra}[1]{\left (#1\right )}
\newcommand{\brac}[1]{\left [#1\right ]}

\newcommand{\pobr}[1]{\left \{#1\right \}}

\newcommand{\e}{\mathcal{E}}
\newcommand{\hk}{\hslash}

\newcommand{\eqreff}[2]{(\ref{#1}-\ref{#2})}

\newcommand{\pmatrx}[1]{\begin{pmatrix} #1 \end{pmatrix}}

\newcommand{\lcases}[1]{\left \{ \begin{aligned} #1 \end{aligned} \right .}

\begin{document}

\markboth{B.M. Szablikowski  and M. B\l aszak}{On deformations of
standard $R$-matrices for integrable systems}

\title{On deformations of
standard $R$-matrices for integrable infinite-dimensional
systems\footnote{Partially supported by KBN research
  grant No. 1 P03B 111 27.}}

\author{B\l a\.zej M. Szablikowski\footnote{E-mail: bszablik@amu.edu.pl }$\ $
and Maciej B\l aszak\footnote{E-mail: blaszakm@amu.edu.pl}\\
Institute of Physics, A. Mickiewicz University,\\ Umultowska 85, 61-614 Pozna\'n, Poland}



\maketitle

\begin{abstract}
Simple deformations, with a parameter $\epsilon$, of classical
$R$-matrices which follow from decomposition of appropriate Lie
algebras, are considered. As a result nonstandard Lax
representations for some well known integrable systems are
presented as well as new integrable evolution equations are
constructed.\\
{\it (To appear in Journal of Mathematical Physics.)} 
\end{abstract}

\section{Introduction}

In the theory of nonlinear evolutionary systems one of the most
important problems is construction of integrable systems. By
integrable systems we understand those which have infinite
hierarchy of commuting symmetries. It is well known that a very
powerful tool, called the classical $R$-matrix formalism, can be
used for systematic construction of field and lattice integrable
dispersive systems (soliton systems) as well as dispersionless
integrable field systems (see \cite{S-T-S}-\cite{BS2} and the
references there).

The crucial point of the formalism is the observation that
integrable systems can be obtained from Lax equations. Let $\alg$
be a Lie algebra, equipped with the Lie bracket $[\cdot, \cdot]$.
A linear map $R:\alg \arrow \alg$, such that the bracket $[a,b]_R
:= [R a, b] + [a,R b]$ is a second Lie product on $\alg$, is
called the classical $R$-matrix. Assume that $R$ satisfies an
Yang-Baxter equation YB($\alpha$): $[R a, R b] - R [a, b]_R +
\alpha [a,b] = 0$, which is a sufficient condition for $R$ to be
an $R$-matrix. Then, powers of $L$ generate mutually commuting
vector fields
\begin{align}\label{evlax}
  L_{t_n} = \brac{R \bra{L^n}, L} .
\end{align}
For fixed $n$ the remaining systems are considered as its
symmetries. In this sense \eqref{evlax} represents a hierarchy of
integrable dynamical systems.

In this article the deformation method for systems \eqref{evlax},
preserving the integrability, is presented. It has been done on
the level of their Lax representations through simple
deformations, with parameter $\epsilon$, of classical
$R$-matrices. It is shown that such a procedure leads to the
construction of nonstandard Lax representations for some well
known integrable systems as well as to the construction of new
integrable evolution equations.

\section{Deformations of standard $R$-matrices}

To construct the simplest $R$-structure let us assume that the Lie
algebra $\alg$ can be split into a direct sum of Lie subalgebras
$\alg_+$ and $\alg_-$, i.e. $\alg =\alg_+ \oplus \alg_-$, where
$[\alg_\pm,\alg_\pm]\subset \alg_\pm$. Denoting the projections
onto these subalgebras by $P_\pm$, we define the $R$-matrix as
\begin{equation}\label{rp}
R = \frac{1}{2} (P_+ - P_-) .
\end{equation}
Straightforward calculation shows that \eqref{rp} solves
YB($\frac{1}{4}$). The classical $R$-matrices constructed in this
way we understand as {\bf standard} ones.

Let us consider the following deformation of \eqref{rp}
\begin{equation}\label{rdef}
R'(a) = R(a) + \epsilon r(a)
\end{equation}
where $\epsilon$ is an arbitrary constant playing the role of a
deformation parameter and $r$ is a linear deformation operator.
First, assume that $r$ satisfies the following two relations
\begin{equation}\label{cond1}
\brac{r a,b_+}\in \alg_+\qquad \brac{r a,b_-}\in \alg_-\qquad
\qquad a\in \alg$, $b_+\in \alg_+$, $b_-\in \alg_- .
\end{equation}
So, the question arises when the deformed $R$ preserves the
property of being $R$-matrix. Once again, straightforward
calculation shows that \eqref{rdef} solves YB($\frac{1}{4}$) when
the following condition is fulfilled
\begin{equation}\label{cond2}
r\brac{a,b}_R + \epsilon r\brac{a,b}_r - \epsilon \brac{r a,r b} =
0
\end{equation}
where $[a,b]_r = [r a, b] + [a, r b]$.

\section{Dispersionless systems}

Let $A$ be the algebra of formal Laurent series (Lax polynomials)
in $p$ \cite{Bl2}
\begin{equation}\label{poalg}
A = \pobr{L = \sum_{i\in \bb{Z}} u_i(x)p^i }
\end{equation}
where the coefficients $u_i(x)$ are smooth functions of $x$. Poisson
brackets on $A$ can be introduced in infinitely many ways as
\begin{equation}\label{pobra}
\brac{f,g}\equiv  \pobr{f, g}_s := p^s \bra{\deriv{f}{p}
\deriv{g}{x} - \deriv{f}{x} \deriv{g}{p}}\qquad s\in \bb{Z} .
\end{equation}
Then, fixing $s$, $A$ is the Poisson algebra with an appropriate
bracket \eqref{pobra}. We construct the standard $R$-matrix,
through a decomposition of $A$ into a direct sum of Lie
subalgebras. For a fixed $s$ let $A_{\me -s+k} = \{\sum_{i\me
-s+k} u_i(x)p^i\}$ and $A_{<-s+k} = \{\sum_{i< -s+k} u_i(x)p^i\}$.
Then, $A_{\me -s+k}, A_{< -s+k}$ are Lie subalgebras in the
following cases:
\begin{enumerate}
\item[1.] $\quad s=0,\ k=0$,
\item[2.] $\quad s\in \bb{Z},\ k=1,2$,
\item[3.] $\quad s=2,\ k=3$.
\end{enumerate}
So, fixing $s$ we fix the Lie algebra structure with $k$ numbering
the standard $R$-matrices given in the following form
\begin{equation}\label{disprmat}
R = \frac{1}{2}(P_{\me -s+k} - P_{< -s+k}) = P_{\me -s+k} -
\frac{1}{2} = \frac{1}{2} - P_{< -s+k}
\end{equation}
where $P$ are appropriate projections. The Lax hierarchy
\eqref{evlax} can be represented by two equivalent representations
\begin{equation}\label{displaxh}
L_{t_q} = \pobr{(L^q)_{\me -s+k},L}_s = -\pobr{(L^q)_{<
-s+k},L}_s.
\end{equation}
Notice that different schemes are interrelated. Under the
transformation
\begin{equation}\label{tran}
  x'=x\quad p'=p^{-1}\quad t'=t
\end{equation}
the Lax hierarchy \eqref{displaxh} defined by $k,s$ and $L$
transforms into the Lax hierarchy \eqref{displaxh} defined by
$k'=3-k,s'=2-s$ and $L'=L$, i.e.
\begin{equation}\label{trans}
L \text{ for } k,\ s \Longleftrightarrow L'=L \text{ for }
k'=3-k,\ s'=2-s.
\end{equation}
In such a situation it is enough to consider the cases of $k=0$ and $k=1$.

We are interested in extracting closed systems for
a finite number of fields. To obtain a consistent Lax equation, the Lax operator $L$
has to form a proper submanifold of the full Poisson algebra $A$, i.e. the left and
right-hand sides of expression \eqref{displaxh} have to coincide. They are given in
the form \cite{BS}
\begin{align}
\label{dk0} s=0,\ k=0\ :&\quad L = p^N + u_{N-2} p^{N-2} + u_{N-3} p^{N-3} + ... + u_1 p + u_0\\
\label{dk1} s\in \bb{Z},\ k=1\ :&\quad L = p^N + u_{N-1} p^{N-1} +
... + u_{1-m} p^{1-m} + u_{-m} p^{-m}
\end{align}
where $u_i$ are dynamical fields. Notice, that powers of $L$, in
general fractional, can be calculated by expanding them around
poles, for \eqref{dk0} around $\infty$ and for \eqref{dk1} around
$\infty$ and $0$. So, for $k=0$ we construct one Lax hierarchy and
for $k=1$ we construct, in general, two mutually commuting Lax
hierarchies.

We are looking for a simple deformation of \eqref{disprmat} in the
form
\begin{equation}\label{form}
r = p^\alpha P_\beta\qquad \qquad P_\beta (L) = \brac{L}_\beta = u_\beta
\end{equation}
which will satisfy \eqref{cond1} and \eqref{cond2} for arbitrary
$\epsilon$. By some straightforward calculations, we find them in
the form
\begin{equation}\label{rd1}
  r = p^{-s+k} P_{k-1} \qquad \text{for} \qquad \lcases{\text{1.}\quad s=0,\ k=0\\
  \text{2.}\quad s=0,\ k=1 \\ \text{3.}\quad s=1,\ k=1}
\end{equation}
and
\begin{equation}\label{rd2}
  r = p^{-s+k-1} P_{k-2} \qquad \text{for} \qquad \lcases{\text{4.}\quad s=1,\ k=2\\
  \text{5.}\quad s=2,\ k=2 \\ \text{6.}\quad s=2,\ k=3}
\end{equation}
We see that deformations of \eqref{disprmat} given by the form
\eqref{rdef},\eqref{form} exist only for distinguish values of $s$
and $k$. Nevertheless, for particular fixed values of $\epsilon$
and fixed $s$ (i.e. fixed Lie algebra) there exist other
deformations of the form \eqref{form}, but they are trivial in the
sense that they relate standard $R$-matrices \eqref{disprmat} with
different $k$. Moreover, deformations \eqref{rd2} are constructed
from \eqref{rd1} by using transformation \eqreff{tran}{trans}.
Hence, the only relevant deformations are \eqref{rd1} and so
further we will consider only them. The deformed $R$-matrices for
the cases in \eqref{rd1} take the form
\begin{equation}\label{rmat}
R' = P_{\me -s+k} - \frac{1}{2} + \epsilon p^{-s+k} P_{k-1} =
\frac{1}{2} - P_{< -s+k} +  \epsilon p^{-s+k} P_{k-1}.
\end{equation}

\paragraph{The case: $s=0$, $k=0$.}\hfill \\

The Lax hierarchy for a deformed $R$-matrix is
\begin{equation}\label{dl1}
L_{t_q} = \pobr{(L^q)_{\me 0}+ \epsilon \brac{L^q}_{-1}, L}_0 =
-\pobr{(L^q)_{< 0} - \epsilon \brac{L^q}_{-1}, L}_0 .
\end{equation}
Consistent Lax equations are obtained for Lax operators of the
form
\begin{equation}\label{lo1}
L = p^N + u_{N-1} p^{N-1} + u_{N-2} p^{N-2} + ... + u_1 p + u_0
\end{equation}
where $u_i$ are dynamical fields. From \eqref{dl1} it follows that
\begin{equation}\label{rel}
\begin{split}
(u_{N-1})_{t_q} &= -\epsilon N ([L^q]_{-1})_x\\
(u_{N-2})_{t_q} &= N ([L^q]_{-1})_x -\epsilon (N-1) u_{N-1}
([L^q]_{-1})_x\\
&\ \ \vdots \ .
\end{split}
\end{equation}
So, for $\epsilon = 0$ the field $u_{N-1}$ becomes
time-independent and without loosing generality we can assume
that it is zero. Then, Lax operator becomes a standard Lax
operator \eqref{dk0} for the case with non-deformed $R$-matrix.
From \eqref{rel} the following relation between $u_{N-1}$ and
$u_{N-2}$ results
\begin{equation}\label{red}
u_{N-2} = -\frac{1}{\epsilon}u_{N-1} + \frac{N-1}{2N} (u_{N-1})^2
\end{equation}
so we can eliminate one of them. Eliminating $u_{N-2}$ we will
consider constrained Lax operator in the form
\begin{equation}\label{laxc1}
L = p^N + u_{N-1} p^{N-1} + \bra{-\frac{1}{\epsilon}u_{N-1} +
\frac{N-1}{2N} (u_{N-1})^2} p^{N-2} + u_{N-3} p^{N-3} + ... + u_1
p + u_0 .
\end{equation}
Reparameterizing \eqref{laxc1}: $u_{N-1} \map -\epsilon u_{N-2}$
and then taking the limit $\epsilon \rightarrow 0$ it becomes the
standard Lax operator \eqref{dk0}.

\begin{lemma}
For arbitrary $\epsilon$ the Lax hierarchy
\eqref{dl1},\eqref{laxc1} is equivalent to the Lax hierarchy
\eqref{displaxh},\eqref{dk0} with $s=k=0$.
\end{lemma}
The sketch of the proof is as follows. We are looking for
transformations that will relate fields from \eqref{laxc1} and
fields from \eqref{dk0}. We postulate the following form of these
relations:
\begin{equation}\label{t1}
   \begin{split}
    u_{N-1}&\map -\epsilon u_{N-2}\\
    u_i&\map u_i + f_i(u_{N-2},u_{N-3},...,u_{i+1})\qquad
    \text{for}\quad N-3\me i\me 0.
   \end{split}
\end{equation}
Then, we construct functions $f_i$ in such a way, that hierarchy
\eqref{dl1} will lead to the same evolution system as
\eqref{displaxh} for $s=k=0$. We compare the first non-trivial
systems from these hierarchies. Functions $f_i$ are recursively
constructed comparing evolution expressions for ${u_{i+1}}_t$.
Such a procedure guarantees that the expressions for the fields
will be the same only for components ${u_{N-2}}_t, ..., {u_1}_t$.
So, the equality between both evolution expressions for $u_0$ has
to be argued. The systems for Lax operators \eqref{dk0} and
\eqref{laxc1} both can be understood as the reduction of
infinite-field systems for Lax operators of the form $L' = a_1p +
a_0 + a_{-1}p^{-1} + ...$, given by constraint $L = L'^N$. The
equivalence between the hierarchies considered, constructed from
$L'$, can be shown by explicit infinite recurrence form
\eqref{t1}. Now, reducing them to finite-field systems one finds
the appropriate transformation between finite-field systems,
including the evolution for $u_0$. So, the Lax hierarchy
\eqref{dl1},\eqref{laxc1} is a new representation of well known
integrable dispersionless hierarchies. The form of transformation
\eqref{t1} , related both systems, guarantees that it is an
invertible transformation.

\begin{example} Dispersionless KdV: $N=2$.

For $L = p^2 + u p +v$ and $L_{t_i} = \{ (L^\frac{i}{2})_{\me 0}+ \epsilon
[L^\frac{i}{2}]_{-1},L\}_0$ we find
\begin{equation}\label{kdv}
\begin{split}
\pmatrx{u\\ v}_{t_1} &= \pmatrx{\frac{1}{2}\epsilon uu_x-\epsilon v_x\\
\frac{1}{4}(\epsilon u-2)(uu_x-2v_x)}\\
\pmatrx{u\\ v}_{t_3} &= \pmatrx{-\frac{3}{16}\epsilon (u^2-4v)(uu_x-2v_x)\\
-\frac{3}{32}\epsilon (\epsilon u-2)(u^2-4v)(uu_x-2v_x)}\\
&\vdots \ .
\end{split}
\end{equation}
In the limit $\epsilon \rightarrow 0$ and $u=0$ it becomes the
standard dispersionless KdV hierarchy. Notice, that fields $u$ and $v$
are not independent. According to \eqref{red}
$v=-\frac{1}{\epsilon}u+\frac{1}{4}u^2$ and the hierarchy \eqref{kdv}
is equivalent to the one
\begin{equation}\label{kdv2}
u_{t_1} = u_x\qquad u_{t_3}=-\frac{3}{2\epsilon}uu_x\quad ...\ .
\end{equation}
i.e. reparameterized dispersionless KdV. The transformation to the
standard form of dispersionless KdV is given by $u\map -\epsilon
u$. The hierarchy \eqref{kdv2} is generated from $L = p^2 + u p
+\frac{1}{4}u^2-\frac{1}{\epsilon}u$.
\end{example}

\begin{example}Dispersionless Boussinesq: $N=3$.

Here we present only the result for the constrained Lax operator
\eqref{laxc1} $L = p^3 + u p^2
+\bra{\frac{1}{3}u^2-\frac{1}{\epsilon}u}p + w$. Then, the first
nontrivial system from the hierarchy is
\begin{align*}
L_{t_2} &= \pobr{ \bra{L^\frac{2}{3}}_{\me 0}+ \epsilon \brac{L^\frac{2}{3}}_{-1},L}_0 \Longleftrightarrow\\
\pmatrx{u\\ w}_{t_2} &=
\pmatrx{\frac{2}{9}(\epsilon u-6)uu_x-2\epsilon w_x\\
\frac{2}{81\epsilon^2}(\epsilon u-3)(\epsilon u(\epsilon u-9)+9)uu_x
-\frac{2}{9}(\epsilon u-6)u w_x} .
\end{align*}
Eliminating the field $w$ we obtain the reparameterized
dispersionless Boussinesq: $u_{tt}  = \frac{2}{3\epsilon}
(u^2)_{xx}$. The transformation \eqref{t1} to the standard form of
the dispersionless Boussinesq system is given by: $u\map -\epsilon
u$, $w\map w -  \frac{1}{3}\epsilon u^2 -\frac{1}{27}\epsilon^3
u^3 $.
\end{example}

\paragraph{The case: $s=0$, $k=1$.}\hfill \\

The Lax hierarchy for the deformed $R$-matrix \eqref{rmat} is
\begin{equation}\label{dl2}
L_{t_q} = \pobr{(L^q)_{\me 1}+ \epsilon  \brac{L^q}_{0} p, L}_0 =
-\pobr{(L^q)_{< 1} - \epsilon \brac{L^q}_{0} p, L}_0 .
\end{equation}
Appropriate Lax operators are of the form
\begin{equation}\label{lo2}
L = u_N p^N + u_{N-1} p^{N-1} + ... + u_{1-m} p^{1-m} + u_{-m}
p^{-m}.
\end{equation}
From \eqref{dl2} one finds that $(u_{N})_t =\epsilon (u_N)_x
[L^q]_{0}- \epsilon N u_N ([L^q]_{0})_x$. Hence, in the limit of
$\epsilon = 0$ the field $u_{N}$ becomes a time-independent field
$c_N$. Fixing $c_N=1$ the Lax operator becomes a standard Lax
operator \eqref{dk1} for $s=0,k=1$. Moreover, there is no
constraint contrary to the previous case. Hence, the Lax hierarchy
\eqref{lo2},\eqref{dl2} leads to new integrable dispersionless
systems, at least to the best of our knowledge. Notice that the
zero power of $L$ always leads to the space translation symmetry:
$L_{t_0} = \epsilon L_x$.

\begin{example}\label{Benney} Extended dispersionless Benney: $N=m=1$.

Let $L = u p + v + w p^{-1}$, then for
$L_{t_i} = \{(L^i)_{\me 1}+ \epsilon [L^i]_{0}p,L\}_0$ we find
\begin{align*}
\pmatrx{u\\ v\\ w}_{t_0} &= \epsilon
\pmatrx{u_x\\ v_x\\ w_x}\\
\pmatrx{u\\ v\\ w}_{t_1} &=
\pmatrx{\epsilon u_xv-\epsilon uv_x\\ uv_x+\epsilon vv_x\\ u_xw+uw_x+\epsilon v_xw+\epsilon vw_x}\\
\pmatrx{u\\ v\\ w}_{t_2} &=
\pmatrx{\epsilon u_xv^2-2\epsilon uvv_x-2\epsilon u^2w_x\\
2uu_xw+2uvv_x+2u^2w_x+\epsilon v^2v_x+2\epsilon uv_xw\\
2u_xvw+2uv_xw+2uvw_x+2\epsilon u_xw^2+2\epsilon vv_xw+\epsilon v^2w_x+4\epsilon uww_x}\\
&\  \vdots \ .
\end{align*}
In the limit $\epsilon \rightarrow 0$ and $u=1$
we obtain the standard dispersionless Benney system.
\end{example}

\begin{example}Two field system: $N=0,m=1$.

Consider $L= v + w p^{-1}$. Then $[L^i]_{0} = v^i$ and $(L^i)_{\me
1}=0$ for $i=0,1,2,...\ $. Hence we obtain the system
\begin{equation}\label{strange}
L_{t_i} = \pobr{\epsilon [L^i]_{0}p,L}_0\quad \Longleftrightarrow
\quad \pmatrx{v\\ w}_{t_i} = \pmatrx{\epsilon v^i v_x\\ \epsilon
\bra{v^i w}_x}
\end{equation}
which does not have any standard counterpart.
\end{example}

\paragraph{The case: $s=1$, $k=1$.}\hfill \\

The Lax hierarchy is
\begin{equation}\label{dl3}
    L_{t_q} = \pobr{(L^q)_{\me 0}+ \epsilon  \brac{L^q}_{0}, L}_1 =
-\pobr{(L^q)_{< 0} - \epsilon  \brac{L^q}_{0}, L}_1
\end{equation}
and appropriate Lax operators take the form
\begin{equation}\label{lax3}
L = u_N p^N + u_{N-1} p^{N-1} + ... + u_{1-m} p^{1-m} + u_{-m}
p^{-m}.
\end{equation}
From \eqref{dl3} it follows that
\begin{equation}\label{r}
(u_{N})_t = -\epsilon N u_N ([L^q]_{0})_x\qquad ... \qquad
(u_{-m})_t = (1+\epsilon) m u_{-m} ([L^q]_{0})_x .
\end{equation}
So, we find that the highest and lowest fields are related by
\begin{equation*}
u_N^{(1+\epsilon)m} = u_{-m}^{-\epsilon N} .
\end{equation*}
For $\epsilon = 0$ the field $u_{N}$ becomes a time-independent
field $c_N$ (let $c_N=1$), then the Lax operator \eqref{lax3}
becomes a standard Lax operator \eqref{dk1} for $s=k=1$. For
$\epsilon = -1$ the field $u_{-m}$ becomes time-independent and
the Lax operator becomes a standard Lax operator for $s=1, k=2$.
This last case follows from the fact that for $s=1,k=1$ and
$\epsilon=-1$ the deformed $R$-matrix \eqref{rmat} becomes the
standard $R$-matrix \eqref{disprmat} for $s=1,k=2$. Eliminating
$u_N$ field the Lax operator takes the form
\begin{equation}\label{laxc3}
L =  u_{-m}^{-\frac{\epsilon N}{(1+\epsilon)m}} p^N + u_{N-1}
p^{N-1} + ... + u_{1-m} p^{1-m} + u_{-m} p^{-m}
\end{equation}
In the limit $\epsilon \rightarrow 0$ it becomes the standard Lax
operator \eqref{dk1} for $s=k=1$.

\begin{lemma}
For arbitrary $\epsilon$, the Lax hierarchy
\eqref{dl3},\eqref{lax3} is equivalent to the Lax hierarchy
\eqref{displaxh},\eqref{dk1} with $s=k=1$.
\end{lemma}
To show this let us make the following transformation
\begin{equation}\label{dt}
u_i\map u_i u_{N}^\frac{i}{N}\qquad p\map
u_{N}^{-\frac{1}{N}}p\qquad \quad \text{for}\quad  N-1\me i
> -m.
\end{equation}
The Poisson bracket \eqref{pobra} for $s=1$ is invariant under
\eqref{dt}. Moreover, the Lax operators \eqref{lax3} transform
into \eqref{dk1} one. Then, after the transformation of
coordinates \eqref{dt} we get
\begin{equation*} L_t\map L_t +
\frac{1}{N} u_{N}^{-1} (u_{N})_t p L_p \overset{\text{by
\eqref{r}}}{=} L_t - \epsilon \bra{\brac{L^q}_{0}}_x p L_p = L_t +
\pobr{\epsilon \brac{L^q}_{0}, L}_1 .
\end{equation*}
Hence, the hierarchy \eqref{dl3} turns into \eqref{displaxh} one
with $s=k=1$.

\begin{example} Extended dispersionless Toda: $N=m=1$.

For Lax operator $L = u p + v + w p^{-1}$ from $L_{t_1} =
\{(L)_{\me 0}+ \epsilon \brac{L}_{0},L\}_0$ we find
\begin{equation*}
\pmatrx{u\\ v\\ w}_{t_1} = \pmatrx{-\epsilon u v_x\\ u_x w + u
w_x\\ (1+\epsilon)v_x w} .
\end{equation*}
In the limit $\epsilon = 0$ and $u=1$ we obtain the standard
dispersionless Toda system. In the limit $\epsilon =-1$ and $w=1$
we obtain the reparameterized dispersionless Toda system. The
transformation \eqref{dt} to the standard case is given by: $v\map
v$, $w\map u^{-1}w$. Eliminating field $u$, for $L =
w^{-\frac{\epsilon}{1+\epsilon}} p + v + w p^{-1}$, we get
\begin{align*}
\pmatrx{v\\ w}_{t_1} = \pmatrx{\frac{1}{1+\epsilon}w^{-\frac{\epsilon}{1+\epsilon}}w_x\\
(1+\epsilon)v_xw} .
\end{align*}
For $\epsilon=0$ or by the transformation: $v\map v$, $w\map
w^{1+\epsilon}$ it becomes the standard dispersionless Toda
system.
\end{example}

Notice that for some dispersionless systems it is possible to
construct their integrable dispersive counterparts: field and
lattice soliton systems. Actually, one can do it on the level of
their Lax representation through Weyl-Moyal like deformation
quantization procedure \cite{BS2} of dispersionless case. The idea
relies on the deformation of the usual multiplication in $A$
\eqref{poalg} to the new associative but non-commutative product
\begin{equation}\label{star}
f\star g = f \exp \bra{\hk p^s \pr_p \otimes \pr_x} g = \sum_{i\me
0} \frac{\hk^i}{i!} \bra{p^s\pr_p}^i f\cdot \pr_x^i g\qquad f,g\in
A
\end{equation}
called $\star$-product. It depends on the formal deformation
parameter $\hk$. The Lie algebra structure is defined by the
commutator $\pobr{f,g}_\star^s = \frac{1}{\hk} \bra{f\star g -
g\star f}$. Then, the $\star$-product \eqref{star} in the limit
$\hk \rightarrow 0$ reduces to the standard multiplication and the
commutator reduces to the Poisson bracket \eqref{pobra} for fixed
$s$. To construct integrable dispersive systems one has to split
the algebra $A$ with the $\star$-product into a direct sum of its
Lie subalgebras and then construct the standard $R$-matrices. It
can be done only for $s=0,1,2$. But, the case $s=2$ is equivalent
to the case $s=0$. The algebra $A$ with $\star$-product
\eqref{star} for $s=0$ is isomorphic to the Lie algebra of
pseudo-differentials operators \eqref{pdalg}, while for $s=1$ is
isomorphic to the Lie algebra of shift operators \eqref{salg} ($\e
= \exp \hk \pr_x$). The first case leads to the construction of
field soliton
 systems, and the second one leads to the construction of lattice soliton  systems.
Obviously, integrable dispersionless systems can be constructed
from integrable dispersive systems in the so-called
quasi-classical (dispersionless) limit: $\pr_t\map \hk \pr_t,
\pr_x\map \hk \pr_x$ and $\hk\arrow 0$.

\section{Field soliton systems}

Let $\alg$ be the algebra of pseudo-differential operators
\cite{OR}
\begin{equation}\label{pdalg}
\alg = \pobr{ L = \sum_{i\in \bb{Z}} u_i(x)\pr_x^i}
\end{equation}
where the multiplication of two such operators uses the
generalized Leibniz rule $\pr^m u = \sum_{s\geqslant 0}
\binom{m}{s} u_{sx} \pr_x^{m-s}$. The Lie algebra structure of
$\alg$ is given by the commutator $[L_1, L_2] = L_1 L_2 - L_2
L_1$. We consider decomposition of $\alg$ in the form $A_{\me k} =
\{ \sum_{i\me k} u_i(x) \pr_x^i\}$ and $A_{< k} = \{L = \sum_{i<
k} u_i(x) \pr_x^i\}$, which are Lie subalgebras for $k = 0,1,2$.
In this cases the standard $R$-matrices are given by
\begin{equation*}
R = \frac{1}{2}(P_{\me k} - P_{< k}) = P_{\me k} - \frac{1}{2} =
\frac{1}{2} - P_{< k}.
\end{equation*}
So, the Lax hierarchy has the form
\begin{equation}\label{pdl}
L_{t_q} = \brac{(L^q)_{\me k}, L} = -\brac{(L^q)_{< k},L} .
\end{equation}
Consistent Lax equations are obtained for Lax operators of the
form \cite{KO}
\begin{align}
\label{slk0}
k=0: &\quad L = \pr_x^N + u_{N-2} \pr_x^{N-2} + ... + u_1  \pr_x + u_0\\
\label{slk1} k=1: &\quad L = \pr_x^N + u_{N-1}\pr_x^{N-1} + ... + u_0 + \pr_x^{-1} u_{-1}\\
\label{slk2} k=2: &\quad L = u_N \pr_x^N + u_{N-1} \pr_x^{N-1} + ... + u_0 + \pr_x^{-1} u_{-1} + \pr_x^{-2} u_{-2} .
\end{align}
where $u_i$ are dynamical fields. Comparing Lax operators
\eqreff{slk0}{slk2} with those for the dispersionless case
\eqreff{dk0}{dk1} for $s=0$ we see that not all dispersionless
systems have dispersive counterparts.

The simple deformations satisfying \eqreff{cond1}{cond2} are the
following ones
\begin{equation*}
  r= P_{k-1}(\cdot ) \pr_x^k\qquad \text{for}\qquad k=0,1\ .
\end{equation*}
Note, that the first case has been considered, in little bit
different manner, earlier in \cite{KSO}. Hence, the deformed
$R$-matrices have the form
\begin{equation*}
R' = P_{\me k} - \frac{1}{2} + \epsilon P_{k-1}(\cdot)\pr_x^k =
\frac{1}{2} - P_{< k} +  \epsilon P_{k-1}(\cdot)\pr_x^k.
\end{equation*}

\paragraph{The case: $k=0$.}\hfill \\

The Lax hierarchy is
\begin{equation}\label{pdlh0}
L_{t_q} = \brac{(L^q)_{\me 0}+ \epsilon  \brac{L^q}_{-1}, L} =
-\brac{(L^q)_{< 0} - \epsilon  \brac{L^q}_{-1}, L}
\end{equation}
and the appropriate Lax operators are given in the form
\begin{equation}\label{lk0}
L = \pr_x^N + u_{N-1}\pr_x^{N-1} +u_{N-2} \pr_x^{N-2} + ... + u_1
\pr_x + u_0.
\end{equation}
From \eqref{pdlh0} one finds
\begin{equation}\label{rel1}
\begin{split}
(u_{N-1})_{t_q} &= -\epsilon N ([L^q]_{-1})_x\\
(u_{N-2})_{t_q} &= N ([L^q]_{-1})_x -\epsilon (N-1) u_{N-1}
([L^n]_{-1})_x
-\epsilon \frac{N(N-1)}{2} ([L^q]_{-1})_{2x}\\
&\ \ \vdots \ .
\end{split}
\end{equation}
Hence, for $\epsilon = 0$ the field $u_{N-1}$ becomes
time-independent $c_{N-1}$ one (let $c_{N-1}=0$), then Lax
operator becomes a standard Lax operator \eqref{slk0}. Expression
\eqref{rel1} implies the relation between fields $u_{N-1},
u_{N-2}$
\begin{equation*}
u_{N-2} = -\frac{1}{\epsilon}u_{N-1} + \frac{N-1}{2N} (u_{N-1})^2
+ \frac{N-1}{2}(u_{N-1})_x.
\end{equation*}
We eliminate the field $u_{N-2}$ and as a result the Lax operators
take the form
\begin{equation}\label{lc1}
L = \pr_x^N + u_{N-1} \pr_x^{N-1} +
\bra{-\frac{1}{\epsilon}u_{N-1} + \frac{N-1}{2N} (u_{N-1})^2 +
\frac{N-1}{2}(u_{N-1})_x} \pr_x^{N-2} +
 ... + u_0 .
\end{equation}
In the dispersionless limit \eqref{lc1} reduces to \eqref{laxc1}.
Reparameterizing \eqref{lc1}: $u_{N-1} \map -\epsilon u_{N-2}$ and
then taking limit $\epsilon \rightarrow 0$ we obtain the standard
Lax operator \eqref{slk0}.

\begin{lemma}
For arbitrary $\epsilon$ the Lax hierarchy
\eqref{pdlh0},\eqref{lc1} is equivalent to the Lax hierarchy
\eqref{pdl},\eqref{slk0}.
\end{lemma}
We are looking for relations between fields from Lax operators
\eqref{lc1} and \eqref{slk0}, respectively. They are given in the
following form:
\begin{equation}\label{t2}
    \begin{split}
    u_{N-1}&\map -\epsilon u_{N-2}\\
    u_i&\map u_i + f_i[u_{N-2},u_{N-3},...,u_{i+1}]\qquad
    \text{for}\quad N-3\me i\me 0.
   \end{split}
\end{equation}
The square brackets in \eqref{t2} mean that functions $f_i$, in
opposite to the case \eqref{t1}, depend not only on $u_i$, but
also on the derivatives $(u_i)_x$,$(u_i)_{xx}$,$...$ . Functions
$f_i$ are constructed in such a way that hierarchy \eqref{pdlh0}
will lead to the same evolution system as hierarchy \eqref{pdl}
for $k=0$. Argumentation that this equality indeed holds is of the
same nature as in Section 3 the paragraph $s=k=0$.

\begin{example} KdV: $N=2$.

For the constrained Lax operator \eqref{lc1} of the form $L =
\pr_x^2 + u \pr_x + \frac{1}{4}u^2 - \frac{1}{\epsilon}u +
\frac{1}{2}u_x$ we find reparameterized KdV:
\begin{equation*}
L_{t_3} = \brac{\bra{L^\frac{3}{2}}_{\me 0} + \epsilon \brac{L^\frac{3}{2}}_{-1},L}
\Longleftrightarrow
u_{t_3}=\frac{1}{4}u_{3x}-\frac{3}{2\epsilon}uu_x.
\end{equation*}
The transformation to the standard form of KdV is given by $u\map
-\epsilon u$.
\end{example}

\begin{example} Deformed Boussinesq: $N=3$.

Let $L = \pr_x^3 + u \pr_x^2
+\bra{\frac{1}{3}u^2-\frac{1}{\epsilon}u + u_x}\pr_x + w$ then
\begin{align*}
L_{t_2} &=  \brac{\bra{L^\frac{2}{3}}_{\me 0} + \epsilon \brac{L^\frac{2}{3}}_{-1},L}
\Longleftrightarrow\\
u_{t_2} &=
-\frac{4}{3}uu_x-u_{2x}+\frac{2}{9}\epsilon u^2u_x+\frac{2}{3}\epsilon u_x^2
+\frac{2}{3}\epsilon uu_{2x}+\frac{2}{3}\epsilon u_{3x}-2\epsilon w_x\\
w_{t_2} &= -\frac{2}{3\epsilon}uu_x+\frac{8}{9\epsilon}u^2u_x+\frac{2}{3\epsilon}u_x^2+\frac{4}{3\epsilon}uu_{2x}
+\frac{2}{3\epsilon}u_{3x}-\frac{8}{27}u^3u_x-\frac{14}{9}uu_x^2\\
&\quad  +\frac{4}{3}uw_x-\frac{10}{9}u^2u_{2x}-\frac{8}{3}u_xu_{2x}+w_{2x}-\frac{14}{9}uu_{3x}-\frac{2}{3}u_{4x}
+\frac{2}{81}\epsilon u^4u_x\\
&\quad +\frac{8}{27}\epsilon u^2u_x^2+\frac{10}{27}\epsilon u_x^3-\frac{2}{9}\epsilon u^2w_x-\frac{2}{3}\epsilon u_xw_x
+\frac{4}{27}\epsilon u^3u_{2x}+\frac{4}{3}\epsilon uu_xu_{2x}\\
&\quad +\frac{2}{3}\epsilon u_{2x}^2-\frac{2}{3}\epsilon uw_{2x}
+\frac{10}{27}\epsilon u^2u_{3x}+\frac{10}{9}\epsilon
u_xu_{3x}-\frac{2}{3}\epsilon w_{3x} +\frac{4}{9}\epsilon
uu_{4x}+\frac{2}{9}\epsilon u_{5x} .
\end{align*}
Eliminating the field $w$ we obtain reparameterized Boussinesq:
$u_{tt}  = (\frac{2}{3\epsilon} u^2 - \frac{1}{3}u_{xx})_{xx}$.
The transformation \eqref{t2} to the standard form of the
Boussinesq system is given by: $u\map -\epsilon u$, $w\map
w-\frac{1}{3}\epsilon u^2-\frac{1}{27}\epsilon^3 u^3
+\frac{1}{3}\epsilon^2 uu_x -\frac{1}{3}\epsilon u_{2x}$.
\end{example}

\paragraph{The case: $k=1$.}\hfill \\

The Lax hierarchy becomes
\begin{equation}\label{pdlh1}
L_{t_q} = \brac{(L^q)_{\me 1}+ \epsilon  \brac{L^q}_{0} \pr_x, L}
= -\brac{(L^q)_{< 1} - \epsilon  \brac{L^q}_{0} \pr_x, L}
\end{equation}
and the appropriate Lax operators have the form
\begin{equation}\label{lk1}
L = u_N\pr_x^N + u_{N-1}\pr_x^{N-1} + ... + u_0 + \pr_x^{-1}
u_{-1}.
\end{equation}
From \eqref{pdlh1} one finds that $(u_{N})_t =\epsilon (u_N)_x
[L^q]_{0}- \epsilon N u_N ([L^q]_{0})_x$. Hence in the limit
$\epsilon = 0$ the field $u_{N}$ becomes a time-independent $c_N$
one, (let $c_N=1$), then Lax operator becomes a standard Lax
operator \eqref{lk1}. There is no constraint contrary to the
previous case. The Lax operators \eqref{pdlh1} with \eqref{lk1}
lead to the construction of new integrable soliton systems, at
least to the best of our knowledge. Again the zero power of $L$
always leads to the space translation symmetry: $L_{t_0} =
\epsilon L_x$.

\begin{example} Extended Kaup-Broer: $N=1$.

Let $L = u \pr_x + v + \pr_x^{-1}w$, then for $L_{t_i} =
[(L^i)_{\me 1}+ \epsilon [L^i]_{0}\pr_x,L]$ we find
\begin{align*}
\pmatrx{u\\ v\\ w}_{t_0} &= \epsilon
\pmatrx{u_x\\ v_x\\ w_x}\\
\pmatrx{u\\ v\\ w}_{t_1} &=
\pmatrx{\epsilon u_xv-\epsilon uv_x\\ uv_x+\epsilon vv_x\\ u_xw+uw_x+\epsilon v_xw+\epsilon vw_x}\\
u_{t_2} &= \epsilon u_xv^2-2\epsilon uvv_x-2\epsilon u^2w_x-\epsilon u^2v_{2x}\\
v_{t_2} &= 2uu_xw+2uvv_x+2u^2w_x+uu_xv_x+u^2v_{2x}+\epsilon v^2v_x+2\epsilon uv_xw+\epsilon uv_x^2\\
w_{t_2} &= 2u_xvw+2uv_xw+2uvw_x-u_x^2w-3uu_xw_x-uu_{2x}w-u^2w_{2x}+2\epsilon u_xw^2\\
&\quad +2\epsilon vv_xw+\epsilon u_xv_xw +\epsilon v^2w_x+4\epsilon uww_x
+\epsilon uv_xw_x+\epsilon uv_{2x}w\\
&\  \vdots \ .
\end{align*}
It is dispersive counterpart of the hierarchy from Example
\ref{Benney}. In the limit $\epsilon \rightarrow 0$ and $u=1$ we
obtain the standard Kaup-Broer system.
\end{example}

\begin{example} Two field system: $N=0$.

For $L= v + w \pr_x^{-1}$ we have $[L^i]_{0} = v^i$ and
$(L^i)_{\me 1}=0$, where $i=0,1,2,...\ $. Then, we obtain for
$L_{t_i} = [\epsilon [L^i]_{0}\pr_x,L]$ again the dispersionless
hierarchy \eqref{strange}.
\end{example}

\section{Lattice soliton systems}

Let $\alg$ be the algebra of shift operators \cite{BM}
\begin{equation}\label{salg}
\alg = \pobr{ L = \sum_{i\in \bb{Z}} u_i(x) \e^i}
\end{equation}
where $\e$ is the shift operator such that $\e^m u(x) = u(x+m)
\e^m$. The Lie algebra structure of $\alg$ is given by the
commutator $[L_1, L_2] = L_1 L_2 - L_2 L_1$. We consider simple
decomposition of $\alg$ in the form $A_{\me k} = \{\sum_{i\me k}
u_i \e^i\}$ and $A_{< k} = \{\sum_{i< k} u_i \e^i\}$, which are
Lie subalgebras for $k = 0,1$. In these cases the standard
$R$-matrix is given by
\begin{equation*}
R = \frac{1}{2}(P_{\me k} - P_{< k}) = P_{\me k} - \frac{1}{2} =
\frac{1}{2} - P_{< k}.
\end{equation*}
The Lax hierarchy is
\begin{equation}\label{lat}
L_{t_q} = \brac{(L^q)_{\me k},L}
 = -\brac{(L^q)_{< k},L} \qquad k=0,1.
\end{equation}
Notice that these two cases are related by simple transformation
$\e\map \e^{-1}$ and $u_i(x-m)\map u_{-i}(x+m)$. Then, $k=0$ goes
to $k=1$ and vice versa. So, it is enough to consider only the
first case. For $k=0$, the appropriate Lax operators are of the
form \cite{BM}
\begin{equation}\label{sl}
 L = \e^N + u_{N-1}(x) \e^{N-1}+ ... + u_{1-m}(x) \e^{1-m} + u_{-m}(x) \e^{-m} .
\end{equation}
The powers of $L$ are in general fractional and can be constructed
in two ways: for $L^\frac{1}{N} = a_1 \e + a_0 + a_{-1} \e^{-1}+
...$ by requiring $(L^\frac{1}{N})^N = L$ and for $L^\frac{1}{m} =
... + a_1 \e + a_0 + a_{-1} \e^{-1}$ by requiring
$(L^\frac{1}{m})^m = L$. Then, in \eqref{lat} we use
$L^\frac{i}{N}$ and $L^\frac{i}{m}$ for $i=0,1,2...\ $. The Lax
hierarchies \eqref{lat} for $k=0,1$ are dispersive counterparts of
the dispersionless hierarchies \eqref{displaxh} for $s=1,k=1$ and
$s=1,k=2$, respectively.

The simple deformations satisfying \eqreff{cond1}{cond2} are of
the form
\begin{equation*}
k=0,1:\qquad r = P_{0}.
\end{equation*}
But for the same reason as above it is enough to consider the case
$k=0$.

\paragraph{The case: $k=0$.}\hfill \\

The deformed $R$-matrix is given by
\begin{equation*}
R = P_{\me 0} - \frac{1}{2} + \epsilon P_0 = \frac{1}{2} - P_{< 0}
+  \epsilon P_0.
\end{equation*}
Hence
\begin{equation}\label{d5}
L_{t_q} = \brac{(L^q)_{\me 0}+ \epsilon \brac{L^q}_0 ,L}
 = -\brac{(L^q)_{< 0}+ \epsilon \brac{L^q}_0 ,L} .
\end{equation}
The appropriate Lax operators are of the form
\begin{equation}\label{dl}
 L = u_N(x) \e^N + u_{N-1}(x) \e^{N-1}+ ... + u_{1-m}(x) \e^{1-m} + u_{-m}(x) \e^{-m} .
\end{equation}
From \eqref{d5} it follows that
\begin{equation}\label{r2}
u_{N}(x)_t = \epsilon u_N(x) \bra{1-E^N}[L^q]_{0}\qquad ... \qquad
u_{-m}(x)_t = (1+\epsilon) u_{-m}(x) \bra{1-E^{-m}} [L^q]_{0}.
\end{equation}
As a result we find that the highest and lowest field are
interrelated in the following way
\begin{equation}\label{rel3}
\bra{\frac{u_N(x)}{u_N(x-m)}}^{1+\epsilon}
=\bra{\frac{u_{-m}(x)}{u_{-m}(x+N)}}^\epsilon  .
\end{equation}
For $\epsilon = 0$ the field $u_{N}$ becomes a time-independent
$c_N$ one, (let $c_N=1$), then the Lax operator \eqref{dl} becomes
a standard Lax operator \eqref{sl} for $k=0$. For $\epsilon = -1$
the field $u_{-m}$ becomes time-independent and the Lax operator
becomes a standard Lax operator for $k=1$. It is so, because for
$k=0$ and $\epsilon=-1$ the deformed $R$-matrix becomes the
standard $R$-matrix for $k=1$.

\begin{lemma}
For arbitrary $\epsilon$, the Lax hierarchy \eqref{d5},\eqref{dl}
is equivalent to the Lax hierarchy \eqref{lat},\eqref{sl} for
$k=0$.
\end{lemma}
Consider the following transformations:
\begin{equation}\label{lt1}
\begin{split}
    {\e'}^N &= u_N(x)\e^N \Longleftrightarrow \e' = a(x)\e\\
    u_i'(x)&=
\begin{cases}
\frac{u_i(x)}{a(x)a(x+1)\cdot...\cdot a(x+i-1)} & \text{for}\quad  N-1\me i> 0\\
u_0(x)& \text{for}\quad  i= 0\\
u_i(x)a(x-1)a(x-2)\cdot...\cdot a(x-i) &\text{for}\quad 0> i
> -m
\end{cases}
\end{split}
\end{equation}
and $t'=t$, where $a(x)$ is given by the following relation
\begin{equation}\label{ra}
u_N(x) = a(x)a(x+1)\cdot...\cdot a(x+N-1) .
\end{equation}
It transforms the Lax operators \eqref{dl} into the Lax operators
\eqref{sl} with $u_i'(x)$ components. From \eqref{lt1} it follows
that
\begin{equation}\label{der}
\bra{\e'^i}_t = \Pi_{i} \bra{\ln a(x)}_t \e'^i
\end{equation}
where
\begin{equation*}
\Pi_i =
\begin{cases}
1 + E + ... + E^{i-1} &\text{for}\quad i\me 1\\
0 &\text{for}\quad i=0\\
-E^{-1} - E^{-2} -  ... - E^{i} &\text{for}\quad i\les -1
\end{cases}.
\end{equation*}
One finds also from \eqref{ra}, that
\begin{equation*}
(\ln a(x))_t = \bra{\Pi_{N}}^{-1}  (\ln u_N(x))_t
\overset{\text{by \eqref{r2}}}{=} \epsilon \bra{\Pi_{N}}^{-1}
\bra{1-E^N}[L^q]_{0}.
\end{equation*}
 Then, using relation $(1-E^N)\Pi_{i} =
(1-E^i)\Pi_{N}$ which is valid for arbitrary $N,i>0$, we have
\begin{align*}
L_t &= L_{t'} + \sum_{i=-m}^N u_i'(x)\bra{{\e'}^i}_t
\overset{\text{by} \eqref{der}}{=} L_{t'} + \epsilon \sum_{i=-m}^N
u_i'(x) \bra{1-E^N} \Pi_{i}   \bra{\Pi_{N}}^{-1}
[L^q]_{0}  \e'^i\\
&= L_{t'} + \epsilon \sum_{i=-m}^N u_i'(x) \bra{1-E^i} [L^q]_{0}
{\e'}^i = L_{t'} + \brac{\epsilon \brac{L^q}_0 ,L}
\end{align*}
where $u'_N(x)=1$. Hence, the hierarchy \eqref{d5} becomes
\eqref{lat} one with $k=0$.

\begin{example} Extended Toda: $k=1$.

For the Lax operator $L = u(x) \e + v(x) + w(x) \e^{-1}$ and
$L_{t_1} = [(L)_{\me 0}+ \epsilon [L]_{0},L]$ we find
\begin{equation}\label{toda}
\pmatrx{u(x)\\ v(x)\\ w(x)}_{t_1} = \pmatrx{ -\epsilon
u(x)\brac{v(x+1)-v(x)}\\  u(x)w(x+1)-u(x-1)w(x)\\
(1+\epsilon)\brac{v(x)-v(x-1)}w(x)} .
\end{equation}
Again, in the limit $\epsilon \rightarrow 0$ and $u(x)=1$ or by
the transformation \eqref{lt1}: $v(x)\map v(x)$, $w(x)\map
\frac{w(x)}{u(x-1)}$ we obtain the standard Toda system. In the
limit $\epsilon =-1$ and $w(x)=1$ we obtain the reparameterized
Toda system. The fields $u(x)$ and $v(x)$ in \eqref{toda}
according to \eqref{rel3} are related by
$u(x-1)^{1+\epsilon}=w(x)^{-\epsilon}$.
\end{example}

\end{document}